\documentclass[conference]{IEEEtran}
\IEEEoverridecommandlockouts
\usepackage[numbers]{natbib}
\usepackage[bookmarks=true]{hyperref}
\usepackage{amsmath,amssymb,amsfonts}
\usepackage{algorithmic}
\usepackage{graphicx}
\usepackage{textcomp}
\usepackage[usenames,dvipsnames,svgnames]{xcolor}
\usepackage{color,soul}
\usepackage{verbatim}
\usepackage[para]{footmisc}
\IEEEsettopmargin{t}{60pt}
\def\BibTeX{{\rm B\kern-.05em{\sc i\kern-.025em b}\kern-.08em
    T\kern-.1667em\lower.7ex\hbox{E}\kern-.125emX}}

\usepackage[colorinlistoftodos,prependcaption,textsize=footnotesize]{todonotes}

\newcommand{\ph}[1]{{\textbf{#1:}}}
\newcommand{\blue}[1]{{\color{blue}#1}}
\newcommand{\added}[1]{{\color{Salmon}#1}}

\title{ACHORD: Communication-Aware Multi-Robot Coordination with Intermittent Connectivity}

\author{
\normalfont
    \parbox{\linewidth}{\centering
    Maira Saboia\textsuperscript{1,2}$^{*}$,
    Lillian Clark\textsuperscript{3}$^{*}$,
    Vivek Thangavelu\textsuperscript{4}$^{*}$, 
    Jeffrey A. Edlund\textsuperscript{1},  
    Kyohei Otsu\textsuperscript{1}, 
    Gustavo J. Correa\textsuperscript{5}, 
    Vivek Shankar Varadharajan\textsuperscript{6},
    Angel Santamaria-Navarro\textsuperscript{7}, 
    Thomas Touma \textsuperscript{8}, 
    Amanda Bouman\textsuperscript{8}, 
    Hovhannes Melikyan \textsuperscript{1}, 
    Torkom Pailevanian\textsuperscript{1},  
    Sung-Kyun Kim\textsuperscript{1}, 
    Avak Archanian \textsuperscript{1},
    Tiago Stegun Vaquero\textsuperscript{1},
    Giovanni Beltrame\textsuperscript{6}, 
    Nils Napp\textsuperscript{4}, 
    Gustavo Pessin\textsuperscript{2}, 
    Ali-akbar Agha-mohammadi\textsuperscript{1}
    }

    \thanks{$^{*}$ denotes equal contribution.}
    \thanks{\textsuperscript{1}Jet Propulsion Laboratory, California Institute of Technology, CA, USA.
    Instituto Tecnológico Vale, MG, Brazil.
    \textsuperscript{3}University of Southern California, CA, USA.
    \textsuperscript{4}Cornell University, NY, USA.
    \textsuperscript{5}University of California, Riverside, CA, USA.
    \textsuperscript{6}Polytechnique Montréal, QC, CA.
    \textsuperscript{7}Institut de Robòtica i Informàtica Industrial, CSIC-UPC, Spain.
    \textsuperscript{8}California Institute of Technology, Pasadena, CA, USA.
    }
    \thanks{©2022. All rights reserved.}
    
}

\begin{document}
\bstctlcite{IEEEexample:BSTcontrol}
\maketitle

\begin{abstract}
    Communication is an important capability for multi-robot exploration because (1) inter-robot communication (comms) improves coverage efficiency and (2) robot-to-base comms improves situational awareness.
    Exploring comms-restricted (e.g., subterranean) environments requires a multi-robot system to tolerate and anticipate intermittent connectivity, and to carefully consider comms requirements, %even when a robot is out of comms range,
    otherwise mission-critical data may be lost.
    In this paper, we describe and analyze ACHORD (Autonomous \& Collaborative High-Bandwidth Operations with Radio Droppables), a multi-layer networking solution which tightly co-designs the network architecture and high-level decision-making for improved comms.
    ACHORD provides bandwidth prioritization and timely and reliable data transfer despite intermittent connectivity. Furthermore, it exposes low-layer networking metrics to the application layer to enable robots to autonomously monitor, map, and extend the network via droppable radios, as well as restore connectivity to improve collaborative exploration. 
    We evaluate our solution with respect to the comms performance in several challenging underground environments including the DARPA SubT Finals competition environment.
    Our findings support the use of % data muling to complement static relay nodes, 
    data stratification and flow control to improve bandwidth-usage.
\end{abstract}

\section{Introduction}
Multi-robot systems can accomplish tasks which are unrealistic for a single robot, especially when it comes to spatially distributed objectives, and allow for redundancy and robustness to individual robot failures~\cite{rossi2018review, dorri2018multi}.
Particularly in harsh environments, supervised autonomy minimizes risks to the individual robots and to the overall mission, but relies on comms with the remote supervisor. 
To mitigate stress on other parts of the solution, it is advantageous to design a multi-robot autonomy solution that maximizes comms capability.

\begin{figure}
\centering
\includegraphics[width=0.9\columnwidth]{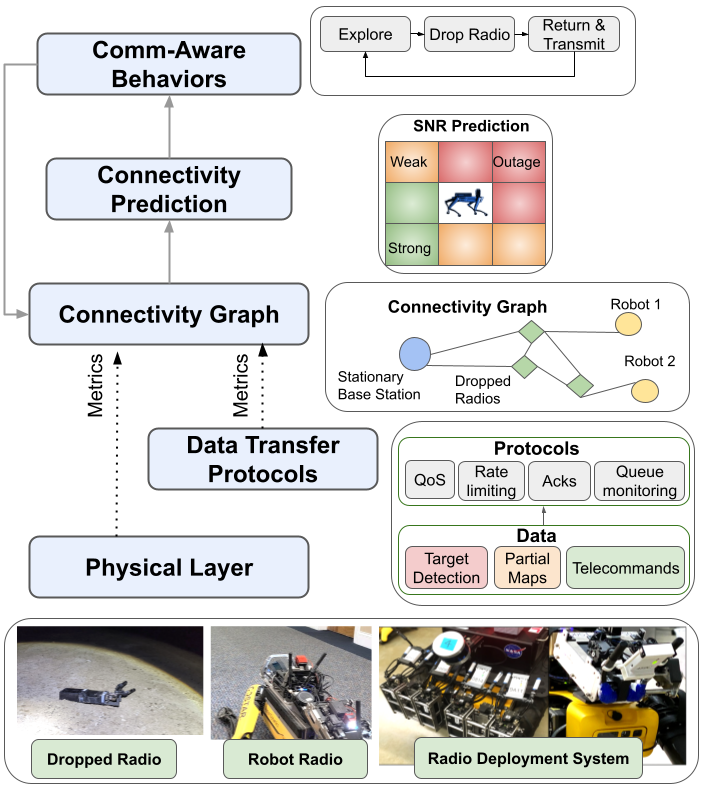}
\caption{ACHORD: At the physical layer, robots use wireless radios (comms nodes) to form a mesh network with droppable radios they deploy. Metrics from the lower-layers of the networking stack (e.g., SNR and queue size) inform high-level connectivity monitoring, prediction, and comms-aware behaviors.}
\vspace{-0.5cm}
\label{fig:achord} 
\end{figure}

% \ph{Problem}
We consider the setting where exploring robots must (1) navigate in an environment where comms might be lost temporarily, (2) communicate findings to a stationary base which requires situational awareness, and (3) share data with nearby robots.
% \ph{Challenges} 
Maintaining comms links is challenging during autonomous exploration of large-scale environments, especially those with winding passages or obstacles which prevent line-of-sight comms.
% For autonomous exploration of large-scale environments, especially environments with winding passages or obstacles which prevent line of sight {\color{blue}communication}, maintaining comm links is challenging. 
Maintaining links at all times constrains the maximum coverage a multi-robot team can achieve.
Even taking approaches to expand the network through relay nodes (static nodes or robots), the effective comms range of the stationary base is limited by the number of nodes. Furthermore, the available bandwidth decreases with each additional network node.
%
% {\color{gray}
% This is challenging because robotic exploration usually generates large amounts of data (e.g. 3D maps of large environments). Transferring pointclouds, images, merged maps, or streaming video in real-time requires high-bandwidth and stable comm links. 
% %
% Efficiently using this bandwidth requires careful management and an understanding of network health (congestion, available bandwidth, interference, etc).
% }
%
An efficient usage of this bandwidth requires careful management and an understanding of the network health (congestion, available bandwidth, interference, etc) considering the large amounts of data typically generated by robotic exploration (e.g., 3D maps of large environments). Transferring pointclouds, maps, or streaming video in real-time requires high-bandwidth and stable comms links.

% \ph{Approach} 
In this paper we address two challenges.
Firstly, we present a comms architecture which meets the needs of exploration: dedicated protocols for distinct classes of data and exposure of statistics on link-quality and queue sizes to upper layers of the networking stack for planning.
Secondly, we present the implementation of behaviors during exploration which meet the needs of comms: robots autonomously determine when to drop additional comms nodes and when to prioritize timely data transfer. 

\subsection{Related Work}

\ph{Comms-aware exploration} 
The communication challenges posed by multi-robot exploration have drawn increased attention \cite{amigoni2017multirobot,guo2018multirobot,klaesson2020planning}. In particular, ensuring the availability of comms links between all robots \cite{stump2008connectivity,robuffo2013passivity,rooker2007multi} or via dedicated relay robots \cite{dixon2009maintaining, yan2012robotic, tekdas2010robotic} are well-researched objectives.
More recently, \textit{intermittent} connectivity has been considered as a more flexible objective, and several works present path-planning methods which ensure intermittent connectivity by requiring robots meet infinitely often \cite{rovina2020asynchronous,kantaros2019temporal,aragues2020intermittent} in a known environment.
During exploration, maintaining connectivity (even intermittently) poses unique advantages as large-scale exploration and connectivity are opposing objectives  \cite{amigoni2017multirobot, clark2021queue}. 
Designing comms-aware exploration strategies requires modeling connectivity, and existing works model link qualities as a function of distance \cite{pei2013connectivity} or other factors like shadowing and multipath components \cite{mostofi2010estimation, clark2022prop}. 
However, decision-making based soley on connectivity fails to consider whether robots have new information to transfer, and realistic memory constraints \cite{guo2018multirobot}. 
In this work, we consider these practical needs and incorporate information about the network state including the size of data buffers at each robot. Recent works have verified this concept theoretically and in simulation 
\cite{clark2021queue,spirin2013time,gasparri2014throughput,klaesson2020planning}, while we consider its implementation in practice as one component of our multi-layer architecture.

% Most of the state-of-the-art research in coordinated multi-robot exploration requires connectivity for all robots at all times.
% %
% Ensuring the availability of comms links is a well-researched objective 
% % \cite{stump2008connectivity,robuffo2013passivity,rooker2007multi,vaquero2018approach}.
% \cite{stump2008connectivity,robuffo2013passivity,rooker2007multi}.
% %
% Previous work has also considered the trade between task-specific goals and increasing connectivity 
% \cite{kantaros2014communication,muralidharan2017path}.
% %
% More recently, approaches to exploration with \textit{intermittent} connectivity have been considered, and a survey of comms-restricted multi-robot exploration is provided in \cite{amigoni2017multirobot}. 
% %
% Most existing works model the multi-robot network as a connectivity graph, where edge weights represent link qualities and are either based on a deterministic comms radius \cite{pei2013connectivity} or a function of distance \cite{stump2008connectivity}.
% %
% Modeling connectivity is informative, but decision-making based on only this information fails to consider whether robots have new information to transfer. 
% In this work, we augment this understanding with information about the network state including the size of data buffers at each robot. Several works have verified this concept theoretically and in simulation 
% % \cite{clark2021queue,spirin2013time,gasparri2014throughput,klaesson2020planning}
% \cite{clark2021queue,gasparri2014throughput,klaesson2020planning}
% , while we consider its implementation in practice.

\ph{Comms protocols}
The most common protocols for data transfer in multi-robot systems are 
% There exist several protocols for data transfer in multi-robot systems, some of the most common being: 
Data Distribution Service (DDS), Robot Operating System (ROS), and Message Queuing Telemetry Transport (MQTT). 
DDS prioritizes performance \cite{Profanter_2019}, while MQTT focuses on a lightweight solution for the Internet of Things and ROS focuses on enabling modular development.
Despite its popularity, previous work has shown that ROS is not well-suited for networks of robots subject to intermittent connectivity \cite{sikand2021robofleet}.
With this in mind, most field-hardened networking approaches rely on custom solutions built directly on UDP/TCP which act as a bridge between ROS-enabled robots \cite{hudson2021heterogeneous, ohradzansky2021multi, otsu2020supervised,tranzatto2021cerberus}.
ROS2 is better-suited for multi-robot systems than ROS and provides significant improvements by leveraging DDS and configurable Quality of Service (QoS) parameters, which allow differentiating between classes of data. However, ROS2 is still missing some key features needed for multi-robot autonomy with intermittent and variable bandwidth connectivity. 
% Issues with ROS2: 1. Lack of dynamic throughput control on a per ip pair basis. (I.e. connections on the same machine should have different throughput limits compared with ethernet or radio links.) Ideally these are responsive to measured available bandwidth. (We used ROS1 for ethernet/same machine and fixed throughput on ROS2/JPL_MM.  This was a partial solution.)
% 2. Smarter data synchronization with optional message ordering (We used JPL_MM)
% 3. Ability to know how much data an agent has to transmit to another agent. (Addressed by DataReporter)
% 4. Per message bandwidth prioritization (Not addressed)
% 5. Protection from inadvertent data transmission. (Used ROS1 bridge to limit topic transfer)

\ph{CHORD}
In our previous work\cite{ginting2021chord}, we introduced CHORD (Collaborative High-Bandwidth Operations with Radio Droppables). CHORD is a hybrid ROS1/2 data transfer solution and demonstrates the advantages of using QoS for radio traffic while using ROS TCP connections for intra-robot communication. However, we observed some issues with network congestion when using ROS2's \textit{reliable} QoS to resend messages after a transmitter rejoins the network. Furthermore, CHORD lacked the necessary bookkeeping to provide queue size monitoring at the application layer. This bookkeeping, introduced in this work in Sec. \ref{sec:protocols}, enables high-level autonomy (described in Sec. \ref{sec:exploration}) which explicitly considers these metrics. Our cross-layer design philosophy allows ACHORD to jointly consider low-level networking and high-level decision-making, while the latter was out of the scope of our previous work.

\begin{comment}
    \ph{CHORD} \added{LC: I'm just putting this here in this level of detail so we all know exactly what was already covered in CHORD.} \cite{ginting2021chord} presents the design and implementation of a multi-robot communication architecture CHORD (Collaborative High-bandwidth Operations with Radio Droppables). The authors discuss the benefits of a hybrid ROS1/2 approach and encourage a wider adoption of ROS2. 
    Real-time topics use \textit{best effort} reliability, \textit{volatile} durability, and \textit{keep last} history policy with queue size 1.
    Reliable topics use a \textit{reliable} reliability policy, \textit{transient local} durability policy, and \textit{keep all} history policy. Results are from Urban.
    Lessons learned: With ROS2 you can avoid having a bookkeeping node (using just QoS) and use built in throttling. An advantage of ROS2 is network isolation. Network monitoring, QoS control, and bandwidth management are important.
    %
    \blue{New things in this work: CHORD didn't have the data reporter which (1) allows us to avoid the network congestion of resending \textit{all} reliable messages and therefore scale to more robots, (2) allows us to selectively reorder messages which reduces delays for some critical topics, (3) exposes end-to-end statistics to the application layer to enable autonomy.
    %
    Outside of the scope CHORD originally: communication monitoring and modeling for decision making, autonomous comm drop, and return to comm.
    }
\end{comment}

\vspace{-0.05cm}
\subsection{Contributions}

We present an overview of ACHORD, as shown in Fig. \ref{fig:achord}, a multi-layer networking solution which focuses on scalability and bandwidth-usage in the joint design of low-level networking and high-level decision-making. We analyze its performance as part of the larger NeBula framework~\cite{agha2021nebula} on a network of up to six robots and up to 13 relay nodes in varied environments.
Contributions unique to ACHORD include:
\begin{itemize}
    % we propose a deployment system to ... based on mobile and statistics nodes.
    % \item \ph{Physical network} The autonomous deployment of mesh network nodes extends the effective comms range of the remote base.
    \item \ph{Bandwidth-aware prioritization} To efficiently use the available bandwidth, we introduce a novel classification of data and leverage dedicated protocols to meet QoS needs.
    \item \ph{Network state representations} To enable comms-aware exploration, we propose a rich representation of the network which considers the radio propagation environment, network congestion, and data queue sizes.
    \item \ph{Predictive signal modeling} To adapt to changes in the dynamic network, we introduce the use of radio propagation models to predict link quality.
    \item \ph{Comms-aware coordination} We propose and implement behaviors such that robots can autonomously prioritize timely data transfer without sacrificing exploration.
\end{itemize}

\begin{figure*}[!ht]
\centering
    \includegraphics[trim={0cm 0.05cm 0.0cm 0.1cm},clip,width=0.9\textwidth]{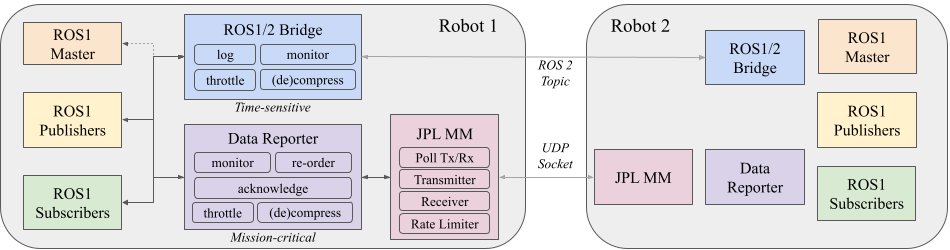}
    \caption{Diagram of software architecture for inter-robot comms. Comms to and from the base is similar.}
    \vspace{-0.5cm}
    \label{fig:protocol_diagram}
\end{figure*}

\section{Problem Formulation}
\label{sec:prob}

We consider a team of heterogeneous robots which explores a large-scale environment in the absence of existing comms infrastructure (e.g., wifi access points) and must communicate findings with a stationary, remote base station.
This operational scenario poses a number of requirements on the system.

\ph{Limited range} First, the range at which robots can communicate directly with the base station is severely limited by the scale of the environment and obstacles or winding passageways which prevent line-of-sight comms.
   This imposes the need for additional comms infrastructure to extend the effective range of the base, for example relay network nodes placed throughout the environment.
   The complexities of the environment also render simple connectivity models insufficient, and more careful monitoring of the network is needed.

\ph{Intermittent connectivity} Second, the exploration objective is to reduce uncertainty by visiting and sensing unexplored areas. Thus, even with the use of relay nodes, it is expected that the robots will explore beyond the effective comms range of the base.
    This imposes the need for a comms architecture which gracefully handles intermittent connectivity. Data transfer protocols should be disruption-tolerant, and reliably transmit all data which is critical to the mission. 
    Further, data transfer protocols should achieve low-latency during periods when connectivity is available, to enable up-to-date situational awareness or teleoperation by human supervisors.

\ph{Limited bandwidth} Third, we assume the multi-robot team generates megabytes of data per minute.
    This imposes the need for efficient use of the limited bandwidth, especially as distance and non-line-of-sight stress link quality.
    After periods without connectivity, robots accumulate large quantities of data and the comms architecture needs to avoid the risk of this flooding the network.

\ph{Network layer metrics} Finally, although the dynamics of the network can be seen as a challenge, the opportunity for controlled mobility presents advantages.
    Leveraging the ability of robots to navigate to areas with connectivity or areas which would benefit from the presence of a dropped relay node imposes the need to circumvent isolation between layers of the network stack. 
    For a cross-layer solution, the comms architecture must expose metrics from the lower layers to the application layer for comms-aware operations.

Given these design requirements, we decompose the high-level problem of ensuring robots can communicate all mission-critical data to the base in a timely manner into layers. The following sections describe the details of ACHORD.

% \ph{IoP}
% \begin{align}
% IoP = f(ns, bw, delay, hops, ...)\\
% ns_k = h(ns_k-1, b_k)\\
% bw_k = h(bw_k-1, b_k)
% \end{align}

% \ph{Problem}
% with abovementioned contrsaitnos, we aim at solvuing the problme of maxinming the infpomraiotn to the operator selecting behaviors that optimally update the network state...
% \begin{align}
%     (ns, b_k) = arg max J(Iop)
% \end{align}

% \ph{challenge}
% oh, sorry,,, this is not computaiotlly tractable... to handle this we propose and dicuss an architecture in the next section that breaks the problem into smaller subproblems and ... 

%%%%%%%%%%%%%%%%%%%%%%%%%%%%%%%%%%%%%%%%%%%%%%%%%%%%%%%%%%%%%%
\section{High-Bandwidth Multi-Robot Networking}
% \section{Exploration-constrained comms} 
% \section{Exploration-centric comms}
% \section{Exploration-oriented comms}
% This section describes the components of our architecture focused on improving comms coverage in environments where large amounts of data are continuously generated and need to be transferred, monitored, and prioritized based on available bandwidth and mission requirements. We discuss the wireless mesh network and present protocols specific to multiple robots with intermittent connectivity.
This section describes the lower layers of our architecture. We discuss the wireless mesh network and present protocols specific to multiple robots with intermittent connectivity. To cover large-scale, comms-limited environments, we create a mesh network using commercially available wireless layer 2 devices (e.g., radios from Rajant or Silvus).

\subsection{Mesh Network Deployment} 
\label{sec:commdrop}
To extend the effective comms range and establish a backbone network, 
robots are equipped with the NeBula Communication Deployment System (NCDS)~\cite{agha2021nebula,funabiki2020range}, shown in Fig. \ref{fig:achord}.
% which allows them to carry additional droppable radios to deploy during the mission to establish a backbone network.
% \textcolor{blue}{
% An updated version of the NCDS was developed for the Final Circuit of the DARPA Subterranean Challenge which improved upon the previous iterations. 
% The NCDS is designed with consideration for modularity across various robotic mobility platforms such as rovers and legged robots, with the ability to carry an arbitrary number of comm nodes, depending on the platform's payload capacity. 
The NCDS has a modular design suited for wheeled and legged robots and can carry up to six droppable radios.
The NCDS can deploy radios upon request from the human supervisor or autonomously via a \textit{scheduler}. 
A serial-ROS connection provides real-time sensory feedback and detailed state logging on which slots have radios loaded and which have deployed.
An important component of the NCDS is automatic jam detection, which allows the scheduler to report a drop failure and initiate a new drop. This improves the resiliency of the system to operational challenges.
% }
% Real-time sensory feedback on which slots have radios loaded and which have actively deployed is provided via a serial-ROS connection, which also supports detailed state logging for debugging. An important addition to the latest design is the automatic jam detection built into the NCDS which enables the scheduler to inform the robot and base station operator if a radio drop has failed. A failed radio drop then triggers a call to the scheduler to initiate a new drop to replace the node that has failed to deploy, improving the resiliency of the system to operational challenges.} 

Each robot monitors the Signal-to-Noise Ratio (SNR)~\cite{schwengler2016wireless} of links between itself and all other radios and uses this information to autonomously deploy the droppable radios to prevent loss of signal. SNR can be measured passively for any received message, and thus does not introduce comms overhead. The Shannon capacity, or theoretical max data capacity for any channel, is given by
% \begin{equation}
% \label{eq:snr}
%     C = B \log_2{(1 + \textrm{SNR})}
% \end{equation}
$C = B \log_2{(1 + \textrm{SNR})}$
where $C$ is the capacity and $B$ is the bandwidth available~\cite{schwengler2016wireless}. This means for a defined bandwidth, increasing SNR increases the amount of data which a link can support. 

% \textcolor{red}{The goal is to keep a lower bound on the bottleneck (minimum along the multi-hop path) SNR value along any route between the robot and the base station.} 
The goal is to keep a lower bound on the \textit{bottleneck SNR} between the robot and base station. The bottleneck SNR of a multi-hop route is the SNR of the weakest link along the route, and we assume data flows along the route with the highest bottleneck SNR.
When this value falls below the desired threshold, the NCDS scheduler is triggered. This typically occurs as the robot gets farther from closest dropped radio, as the backbone network links are above the lower bound by design. If the bottleneck SNR falls below the lower bound, the robot is able to backtrack along its path before committing to the drop location. Our solution then evaluates the traversibility risk~\cite{fan2021step}, the terrain inclination, and the environment geometry~\cite{vaquero2020traversability} to locally select the exact drop position, favoring junctions, dry, and flat surfaces.

\subsection{Data Transfer Protocols}
\label{sec:protocols}

This section describes our inter-robot comms software architecture which meets the requirements enumerated in Sec. \ref{sec:prob}, as shown in Fig. \ref{fig:protocol_diagram}.
% \ph{ROS1} 
For intra-robot comms, we leverage ROS for ease of development. Using ROS for intra-robot comms also allows us to use high bandwidth links on the robot without the limitations of ROS2's throughput controller and isolates the radio traffic to only topics explicitly shared with ROS2 or JPL MM.

\ph{Data classes} For inter-robot comms, we consider three types of data to transfer: (i) key; (ii) mission-critical; and (iii) time-sensitive.
(i) Key data refers to information that needs to be shared periodically and in-order, which is required for the nominal multi-robot mission control, but has no timing restrictions. 
Examples of key data are the telemetry of the robots or the incremental maps they share.
(ii) Mission-critical data includes crucial asynchronous information. While we desire low-latency, the correct transfer of this information is of higher priority than its transmission time.
An example of mission-critical data is the detection and localization of a target object.
(iii) 
Time-sensitive data is selected considering potential harms or vehicle integrity risks, thus an example includes sharing relative positioning between neighboring robots in a collision trajectory.
% \ph{Compression} 
All data which is transmitted over the wireless network is compressed
into a generic, compressed data blob using bzip2\footnote{https://www.sourceware.org/bzip2/}. 

% "We also performed algorithmic data fragmentation if possible. For example, a robot incrementally sends part of the map to be later reassembled to a full map on the base station." -CHORD

\ph{ROS2/DDS} For neighbor discovery and transmitting time-sensitive data, we use ROS2 DDS (eProsima FastRTPS). As described in \cite{ginting2021chord}, each robot has a ROS1/2 bridge, based on the ros1\_bridge package \footnote{https://github.com/ros2/ros1\_bridge}.
% \ph{DDS shortcomings} 
While ROS2/DDS offers many improvements specific to multi-robot networking, it has two shortcomings that we did not address in CHORD \cite{ginting2021chord}: (1) it lacks the option to resend only certain messages after reconnecting with the network, limiting the developer to select a static number of messages to resend (if too large, this will flood the network), and (2) it does not offer the application-layer a way to monitor the amount of data waiting to transmit, which is desirable in our case for high-level decision-making. For disruption-tolerant networks, a resend policy at the application layer which also offers buffer size monitoring is key. 
% 1. Lack of dynamic throughput control on a per ip pair basis. (I.e. connections on the same machine should have different throughput limits compared with ethernet or radio links.) Ideally these are responsive to measured available bandwidth. (We used ROS1 for ethernet/same machine and fixed throughput on ROS2/JPL_MM.  This was a partial solution.)
% 2. Smarter data synchronization with optional message ordering (We used JPL_MM)
% 3. Ability to know how much data an agent has to transmit to another agent. (Addressed by DataReporter)
% 4. Per message bandwidth prioritization (Not addressed)
% 5. Protection from inadvertent data transmission. (Used ROS1 bridge to limit topic transfer)

\ph{Data Reporter} 
To address this, we introduce the \textit{Data Reporter}, which monitors reliable data transfer. For each reliable (key or mission-critical) message sent over the network, an acknowledgement message (ACK) is sent from the receiver back to the sender. On the sender side, messages from data publishers wait in per-topic queues (buffers) to be transmitted and are removed from the queue when an ACK is received. On the receiver side, per-topic queues hold received messages which are then published via ROS to data subscribers. The receive side queue implements re-ordering for key data.
Because the data reporter is implemented in ROS, metrics on per-topic queue sizes can easily propagate up to higher levels of ACHORD as shown in Fig. \ref{fig:achord}.

\ph{JPL MM} 
Key and mission-critical topics require a solution which is highly configurable and provides guarantees of message delivery.
JPL multi-master (JPL~MM) \cite{otsu2020supervised} is a module which provides inter-robot comms that is compatible with ROS and built on top of the User Datagram Protocol (UDP). JPL~MM allows specifying configurations for each ROS topic and network port, and provides many customizable features.
The primary responsibility of JPL~MM on the sender side is to chunk the data into UDP datagrams and transmit them. Then on the receiver side data is reassembled into messages. JPL~MM provides Selective Repeat ARQ (automatic repeat-request) using datagram-ACKs to ensure reliability at the transport layer. It also supports additional compression as needed.
One of the main advantages of JPL~MM for this application is that it provides token bucket rate limiting. A fixed number of tokens is allocated representing the maximum bandwidth available, and these tokens are allocated to each key/mission-critical ROS topic. This lets us prevent certain types of messages from overwhelming the network.
%
% Another advantage is that since JPL~MM is implemented within ROS, it takes advantage of the available ROS logging tools.

%%%%%%%%%%%%%%%%%%%%%%%%%%%%%%%%%%%%%%%%%%%%

\section{Comms-aware multi-robot exploration}
\label{sec:exploration}
% This section describes components of the autonomous exploration which explicitly model and prioritize connectivity while maximizing coverage of unknown environments.
This section describes the higher layers of our architecture. We discuss modeling and prioritizing connectivity while maximizing coverage of unknown environments.

\subsection{State Representation}
\label{sec:state_representation}
We introduce two representations of state used for comms-aware decision-making: the spatial environment state and network performance state.

\begin{figure}
\centering
\includegraphics[trim={0cm 0.cm 0.2cm 0.80cm},clip,width=0.4\textwidth]{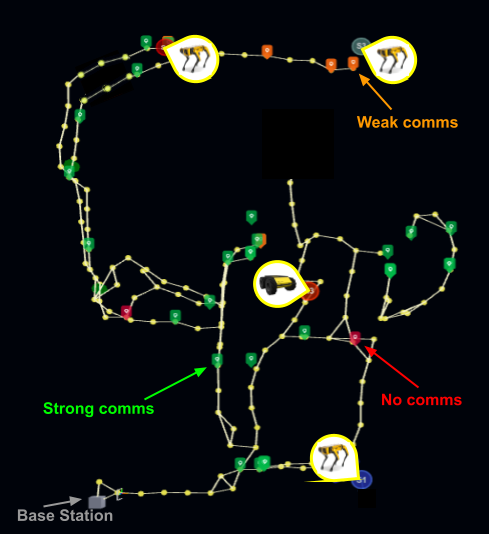}
% \caption{Information RoadMap (IRM) for environment state representation. Comm checkpoints with green, orange, and red tag markers represent high, weak, and none comm levels, respectively.}
\caption{Information RoadMap (IRM) constructed collaboratively during exploration of the DARPA SubT Finals competition environment. The environment state is indicated by green, orange, and red comms checkpoints which represent strong, weak, and no comms, respectively.}
\vspace{-0.5cm}
\label{fig:environmen_representation}
\end{figure}

\ph{Spatial environment state} 
We maintain a semantic, spatial representation of the environment called an Information RoadMap (IRM)~\cite{kim2021plgrim}. As illustrated in Fig. \ref{fig:environmen_representation}, an IRM is a generic graph that captures the environment via nodes and edges, where nodes represent locations and edges connect nodes if a robot can travel between them. We categorize IRM nodes into four types: i) frontier nodes in unexplored space, ii) breadcrumb nodes in previously visited locations, iii) comms checkpoints, which are nodes associated with a signal strength, and iv) dropped radio nodes. Thus, the backbone network topology (as illustrated in Fig. \ref{fig:achord}) is captured by the IRM. It is an incrementally built, shared structure such that when robots meet or return to the base they merge their IRMS, favoring more recent data.

% \ph{Comm checkpoint}
Each comms checkpoint stores the bottleneck SNR value a robot would experience at that location. Comms checkpoints with SNR $\ge T_C = 20$dB are considered strong (green tag markers in Fig.~\ref{fig:environmen_representation}), while checkpoints with $0 \ge$ SNR $< T_C$ are considered weak (orange tag markers).
%
% For weak signal strength, i.e., $0 \ge SNR < T_C$, the comms checkpoints are depicted with orange tag markers.
When SNR $= 0$ (red tag markers), we nominally do not have comms, although the network may transmit some packets sporadically.

\ph{Network performance state} 
The spatial comms representation is unaware of the network usage; a location with a high SNR may exhibit slow data transfer rates due to congestion or interference. For effective comms-aware operations, the robots also maintain statistics on the reliable data that needs to be transferred to the base or other robots: (1) buffer size: the amount of data (in bytes) that needs to be transferred; (2) measured data rate: the amount of data transferred per unit time; and (3) estimated transfer time: the amount of time required to empty the buffer. These properties together determine the network state and help autonomy to prioritize exploration over connectivity or vice versa.

\begin{figure}
    \centering
    \includegraphics[trim={0cm 0cm 0cm 0cm},clip, width=0.8\columnwidth]{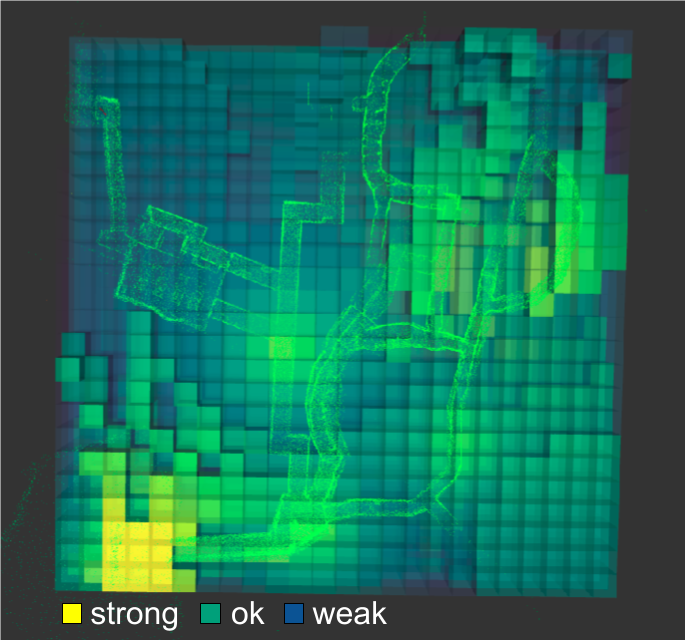}
    \caption{Connectivity map of the DARPA SubT Finals competition environment based on predicted signal strength, which can inform autonomy and augment situational awareness for the human supervisor. 
    % \blue{This map corresponds to the IRM illustrated in Figure~\ref{fig:environmen_representation}.}
    }
    \vspace{-0.5cm}
    \label{fig:ssm_grid}
\end{figure}

\subsection{Predictive Comm Modeling}
Our network topology is dynamic by design, given the mobility of the robots and the deployment of relay nodes. This introduces a challenge for relying on the environment state, as comms checkpoints can become outdated. To overcome this, our solution uses a signal propagation model to update the spatial environment representation with predicted SNR values. We assume the noise level $\sigma_{dB}$ at a comms checkpoint location is invariant and focus on predicting signal strength. 

\ph{Signal propagation model}
Radio signal propagation is a multi-scale process,
but first-order models typically calculate the free space path loss,
which quantifies the expected attenuation in an obstacle-free environment \cite{schwengler2016wireless}. Path loss is modeled as a logarithmic function of distance $d$ given by
\begin{equation}
\label{eq:simple_pathloss}
    PL_{dB}(d) = PL(d_0)_{dB} + \eta 10 \log_{10} (d / d_0).
\end{equation}
$PL(d_0)_{dB}$ is the reference path loss in dB at a known distance $d_0$, and $\eta$ is the path loss exponent which captures how quickly the signal falls off.
Values $\eta = 3.83$ and $PL(d_0 = \textrm{1m})_{dB} = 34$ were selected after fitting this model to experimental data via linear regression, giving the predicted SNR received by $j$ from transmitter $i$:
% \begin{equation}
% \label{eq:snr_prediction}
%     \textrm{SNR}_{dB} = \max_{\textrm{node } i}{(\textrm{Tx}_i - PL_{dB}(d_i) - \sigma_{dB})}.
% \end{equation}
\begin{equation}
\label{eq:snr_prediction}
    \textrm{SNR}_{dB}(i,j) = \textrm{Tx}_{dB}(i) - PL_{dB}(d(i,j)) - \sigma_{dB}(j)
\end{equation}
$\textrm{Tx}_{dB}(i)$ is the transmit power and $d(i,j)$ is the distance between $i$ and $j$.
In a previous work, we have shown that more accurate prediction can be achieved by modeling second and third order effects of the environment, at the cost of increased complexity \cite{clark2022prop}.
% While more accurate prediction can be achieved by modeling the effect of the environment on signal strength, we found this simple model to be sufficient in practice.

\ph{Connectivity maps} Beyond updating checkpoints, having a predictive comms model allows estimating the connectivity offered at arbitrary locations based on the position estimates of the relay nodes, as illustrated in Fig. \ref{fig:ssm_grid}. As depicted in Fig. \ref{fig:achord}, the information in this predicted map flows up to the comms-aware behaviors discussed in the next section.

%%%%%%%%%%%%%%%%%%%%%%%%%%%%%%%%%%%%%%%%%%%%%%%%%%%

%%%%%%%%%%%%%%%%%%%%%%%%%%%%%%%%%%%%%%%%%%%%%%%%%%%
\subsection{Comms-Aware Coordination}
Equipped with the metrics exposed by our low-layer protocols and the state representation described in Sec. \ref{sec:state_representation}, robots autonomously make comms-aware decisions. Here we describe several elements of coordinated comms-aware operations.

\ph{Return to Comms}
When the buffer sizes exceeds an upper bound ($T^u_{B} = 300$KB), the robot will sacrifice nominal exploration and instead move towards an area with strong comms. Our solution selects the closest comms checkpoint from the IRM with SNR $\ge T_C$, or a frontier neighboring this strong comms checkpoint, and moves towards it. If the buffer size drops below a desired threshold ($T^l_{B} = 200$KB) before reaching the target checkpoint, nominal exploration continues immediately. Otherwise, the robot will wait at the target checkpoint until the buffer size drops below $T^l_{B}$. If the buffer size does not decrease within a timeout (60 secs), which could indicate that the network is congested, the robot will choose a strong comms checkpoint closer to the base station as the new target. It will continue in this manner until the buffer size drops.

\ph{Radio deployment coordination} Coordination between robots is required during radio deployment to prevent redundant deployments in the same area. To enable this coordination, robots communicate their intentions via the shared IRM. When a robot reaches the triggering condition to deploy a radio, as described in Sec.~\ref{sec:commdrop}, the robot includes a dedicated node in the IRM that represents the intention to deploy a radio at that position. Before deploying an additional radio, other robots will consider whether the expected radio coverage of the two dropped radios would have significant overlap, and skip redundant deployments.

% \ph{Data muling and rendezvous}
% Since robots share map information with each other, they can act as data mules~\cite{klaesson2020planning}. For example, if two robots are both out of comms range with the remote base but synchronize their maps, and only one returns to connectivity, it can share all map data. This rendezvous behavior can be opportunistic, or guided by waypoints from the human supervisor~\cite{agha2021nebula}.%In the future, we will consider autonomous rendezvous based on the environment and network states.

% KY %%%%%%%%%%%%%%%%

\begin{figure}
    \centering
    \includegraphics[trim={1.0cm 0.1cm 0.0cm 1.2cm},clip,width=0.95\columnwidth]{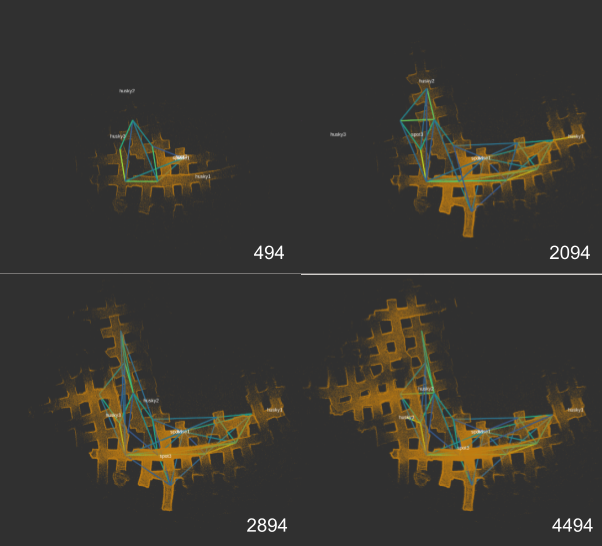}
    \caption{Snapshots of the network during exploring of KY Underground.  Colors indicate the signal strength of links (yellow is strong, purple is weak). Timestamps (sec) are given in the lower right corner of each snapshot.}
    % \added{Sung: 
    % I guess the edge color is visualizing the comms strength, which is important in this figure, but it is not very readable in the current quality. 
    % It would be also good to have similar snapshots before/after comms drop, before/after RTC. (In the latter case, it would be great to visualize the amount of data in the queue.)}}
    \label{fig:ky_snapshots}
\end{figure}

% \begin{figure}[htpb]
%     \centering
%     \includegraphics[width=\columnwidth]{images/sep18d/summary_outages.png}
%     \caption{KY Underground delay/connectivity outages
%     \added{Sung: It would be good to mark when we drop comms, when we start/exit RTC in the timeline. Minor: font is too small.}}
%     \label{fig:ky_delay}
% \end{figure}

% Comp2 %%%%%%%%%%%%%%%%

\begin{figure}
    \centering
    \includegraphics[trim={1.2cm 0.1cm 0.0cm 1.5cm},clip,width=0.95\columnwidth]{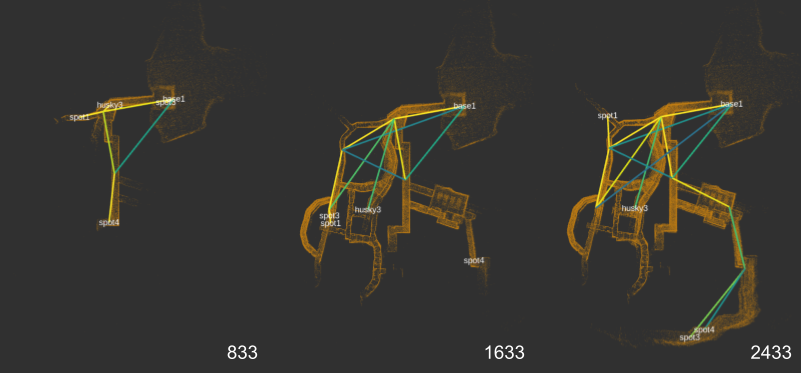}
    \caption{Snapshots of the network during exploration of the DARPA SubT Finals competition environment, day 2.}
    \vspace{-0.5cm}
    \label{fig:comp2_snapshats}
\end{figure}

% \begin{figure}
%     \centering
%     \includegraphics[width=\columnwidth]{images/comp2/delays.png}
%     \caption{Comp 2 delay/connectivity outages. Here we see that spot 4 was out of range for a long time and built up a significant delay. Spot 4 actually failed, but Spot 3 was able to receive and relay its data through a node it dropped. \added{Sung: It would be good to mark when we drop comms, when we start/exit RTC in the timeline. 
%     % Minor: font is too small.
%     }}
%     \label{fig:ky_delay}
% \end{figure}

% \begin{figure}[htpb]
%     \centering
%     \includegraphics[width=\columnwidth]{images/comp3/data_rates.png}
%     \caption{Comp2 data rates (kbps).}
%     \label{fig:ky_rates}
% \end{figure}

\section{Experimental Results}
This section evaluates ACHORD in several comms-restricted environments and discusses lessons learned.

% \ph{Experimental setup}
Our robot team included three Boston Dynamics Spot robots and three Clearpath Robotics Husky robots.
The field test took place in a limestone mine (KY Underground) depicted in Fig. \ref{fig:ky_snapshots}. The mine spanned hundreds of meters, with thick (20m) columns preventing line-of-sight comms.
The other environment we tested in was constructed for the DARPA Subterranean Challenge depicted in Fig. \ref{fig:comp2_snapshats}. The DARPA SubT Finals course was characterized by narrow, winding passageways and had three distinct subsections: an urban environment similar to a subway station, a mine-like environment, and a subterranean cave-like environment. 
% On the first and third days of the competition, all six robots explored. On the second day of the competition, only four robots were deployed.

% \ph{Silvus Radios} 
Given our priority of high-bandwidth comms, ACHORD leverages commercial off-the-shelf MIMO radios from Silvus Technologies (Streamcaster 4240 for the robots and 4400 for the base station) which are designed for mobile ad-hoc networks. The Silvus radios offer multi-hop routing and layer 2 protocols, and equipping the robots with these mesh nodes enables them to act as relays or data mules as needed.
Silvus also provides %a web UI for monitoring the network configuration and link qualities, a 
an API for collecting link quality metrics like SNR, loss rate, noise level, etc. This allows us to propagate these metrics from the physical and link layer up to the application layer.

\begin{figure}
    \centering
    \includegraphics[trim={0.1cm 0.4cm 0.2cm 0.3cm},clip,width=\columnwidth]{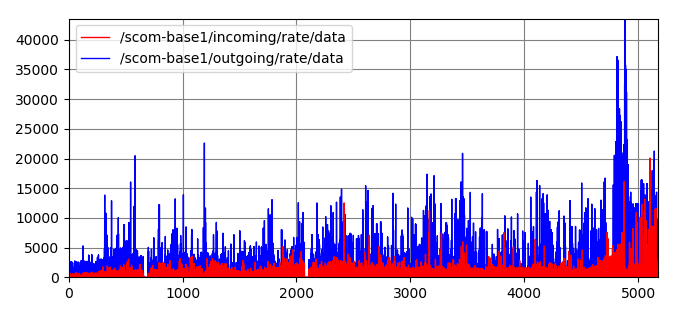}
    \caption{KY Underground data rates (kbps) received by the base station (red) and transmitted by the base station (blue). Robots actively explore from 1019 to 4609 sec. After exploration ended, as the robots returned to the base, the data rates increased.}
    \label{fig:ky_rates}
\end{figure}

% \ph{Results} 
Table~\ref{tab:results} summarizes results on the overall performance of our system with respect to its ability to enable high-bandwidth comms and give the human supervisor a thorough understanding of the environment. The number of dropped radios indicates how much network infrastructure the robots were able to autonomously deploy, which also depends on the scale of the environment. The much larger field test environment required more deployed nodes. The maximum delay indicates the greatest period of time for which any robot was not able to transfer data to the base station. Note that exploration beyond the range of the deployed infrastructure, which is desirable, requires outages.
The effective comms range is a measure of the longest distance from the base any robot was able to travel without losing connectivity through one or more hops. 
Up time captures the maximum percentage of time an exploring robot is able to maintain connectivity with the base through one or more hops, as indicated by low data reporter buffer sizes.
The data rates at the base station, for both incoming and outgoing traffic, are an indicator of overall bandwidth the network is able to support (see Fig. \ref{fig:ky_rates}. Traffic from the base includes primarily mission status updates and the aggregated mapping data.
The key takeaways are that ACHORD enables comms over more than 150 meters of exploration and our network can support up to 20Mbps of data from six exploring robots.

\begin{table}
    \centering
    \caption{Summary of Results}
    \begin{tabular}{c|c|c|c|c}
         & Field Test & Day 1 & Day 2 & Day 3 \\
        \hline \hline
        Time exploring (mins) & 60 & 30 & 30 & 60 \\
        \hline
        Exploring robots (\#) & 6 & 6 & 4 & 6 \\
        \hline
        Deployed radios (\#) & 13 & 1 & 6 & 7 \\
        \hline
        Maximum delay (sec) & 813 & 49 & 1058 & 98 \\
        \hline
        Effective comm range (m) & 173 & 70 & 86 & 68 \\
        \hline
        % \added{\# Critical Messages Dropped} & & & & \\     
        % \hline
        Up time (mins) & 16 & 14 & 20 & 47\\  
        \hline
        % \added{coverage ($\%$)} & & & & \\
        % \hline
        % Max hops (\#) & & & & \\
        % \hline
        Peak data rate & 22.61 & 19.60 & 34.58 & 19.39 \\
        Base $\rightarrow$ Robots (Mbps) & & & & \\
        \hline        
        Peak data rate & 12.43 & 14.41 & 17.14 & 20.30 \\
        Robots $\rightarrow$ Base (Mbps) & & & & \\
    \end{tabular}
    \vspace{-0.0cm}
    \label{tab:results}
\end{table}

\begin{figure}
    \centering

    \includegraphics[trim={0.1cm 0.0cm 1.0cm 0.0cm},clip, width=0.96\columnwidth]{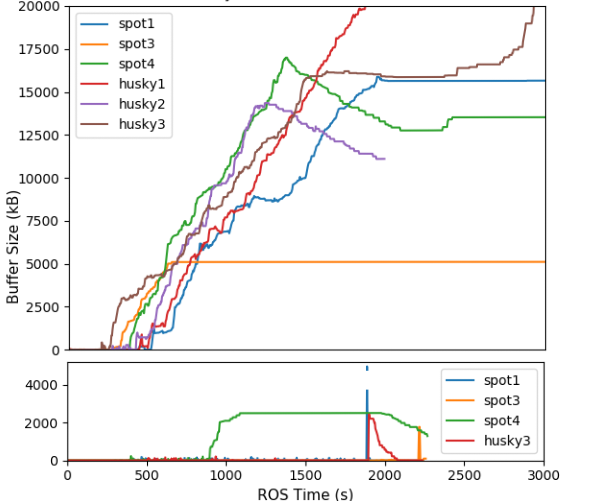}

    \caption{Size of the data reporter buffer (summed over all key and mission-critical topics). The top graph depicts results from Day 1 with the data classification from CHORD and the bottom graph depicts results from Day 2 with the new data classification.
    % represents setting (six robots and one dropped radio) and Day 2 (four robots and six dropped radios) of exploration.
    }
    \vspace{-0.5cm}
    \label{fig:combined_datareporter}
\end{figure}

% \subsection{Discussion}

% \ph{Performance} 
\ph{Comparison with CHORD}
While the many novel aspects of ACHORD make a direct comparison to CHORD \cite{ginting2021chord} challenging, we can highlight the performance improvement offered by certain features.
For example, Fig.~\ref{fig:combined_datareporter} shows the sizes of data reporter queues at each robot during exploration for two settings of ACHORD.
In the first setting (top graph), we required ordered receipt for mission-critical topics, as in CHORD. We observed that the robots built up large queues, and even after restoring connectivity (e.g., spot4 at ~ 1400s), we observed congestion likely due to unnecessary retransmissions which strained the available bandwidth. In the second setting (bottom graph), we introduced the additional data stratification presented in Sec. \ref{sec:protocols} and observed significantly less build up.
% On Day 1, with only 1 dropped radio, the robots built up large queues. Even after restoring connectivity (e.g., spot4 around 1400s), we observed congestion. This was because, at the time, we required mission-critical topics to be ordered. If any messages were dropped, which was likely given the low-SNR links, subsequent messages were retransmitted unnecessarily. This further strained the available bandwidth. On Day 2, after modifying the architecture such that mission-critical messages were allowed out-of-order, we did not observe this congestion. Further, increasing the number of dropped radios to six meant that only one robot, Spot 4, experienced substantial outages. 
% All other buffers remained low.

\ph{Lessons learned} 
We observed a scenario in which a robot got stuck while out of comms range and another robot acted as a data relay, recovering data which would have otherwise been lost. This capability proved a significant advantage, and leads us to conclude that equipping high-level decision-making with inter-robot comms models is key.
% We really benefited from the fact that robots could share data with each other and act as data mules and/or mobile relay nodes. With that in mind, allowing the autonomy to also consider signal strength to neighboring nodes during decision-making could be a direction for future development.
%
With this in mind, high-fidelity simulation of the wireless network would have aided in development and testing of comms-aware autonomy.
In smaller environments, we found it was sufficient to drop nodes generously. The criteria for autonomous node deployment needs careful consideration in expansive environments where droppable nodes are a relatively scarce resource.
If the network infrastructure covers enough of the environment, we found it was sufficient to return to comms sparingly.
With sophisticated autonomy that can reliably determine when to return and transfer data, a small effective comms range and long outages are permissible. On the other hand, with a large effective comms range, a human supervisor has better opportunities to control and intervene, and less sophisticated autonomy is permissible. To design a resilient multi-robot system, we found it is not enough to focus on only one of these objectives, and considering both connectivity and autonomy is key.
% On the other hand, in multi-robot scenarios, sophisticated multi-robot autonomy is tightly associated with inter-robot comms for coordination. Where a system design sits in this trade space should depend on the needs of the application.
%

\section{Conclusion}
In this work, we present an overview of ACHORD, our multi-layer networking solution which provides timely and reliable data transfer for intermittently connected multi-robot teams and leverages droppable radios and comms-aware operations to improve connectivity.
ACHORD is field-hardened through experiments in several underground environments with teams of up to six robots. Our finding indicate that taking the radio environment and network state into account is advantageous for multi-robot exploration when a remote base needs to be kept situationally aware. Autonomous relay node deployment extends the effective comms range of the system, improves signal quality, and reduces delays and connectivity outages. Data prioritization and efficient bandwidth usage are key to enabling exploration of large-scale environments with multiple robots.
Better handling of data prioritization at the semantic level and improved strategies for controlling access to the shared wireless medium are two directions for further study.

% \ph{Future work} 
% In the future we would also like to address better handling of data prioritization at the semantic level, which requires a cross-layer solution.
% %
% We also noted that when many exploring robots return to connectivity simultaneously, congestion could be a problem despite token-bucket flow control. Incorporating strategies to control access of the shared wireless medium, especially at the application layer where it may be possible to also consider semantic prioritization, is an interesting direction for further study.

\section*{Acknowledgements}

We gratefully acknowledge all members of Team CoSTAR. We also thank Belal Wang and Silvus Technologies for their hardware and technical expertise, and Kentucky Underground Storage for access to their facilities. This research was carried out at the Jet Propulsion Laboratory, California Institute of Technology, under a contract with the National Aeronautics and Space Administration (80NM0018D0004). This work was supported in part by FUNDEP, by NASA Space Technology Research Fellowship Grant No. 80NSSC19K1189, and by NSF Grant No. \#1846340.

\bibliographystyle{IEEEtran}
\bibliography{main}

\end{document}